\documentclass[11pt]{article}
\usepackage{graphicx}
\usepackage{epstopdf}
\usepackage{amsfonts,amsmath,amssymb}
\begin{document}

\title{Covariant extrinsic gravity and the geometric origin of dark energy}
\author{S. Jalalzadeh\thanks{s-jalalzadeh@sbu.ac.ir}\,\,\, and\,\, T. Rostami \\
\small Department of Physics, Shahid Beheshti University G. C., Evin, Tehran 19839, Iran}
\maketitle
\begin{abstract}
We construct the covariant or model independent induced  Einstein-Yang-Mills field equations on a 4-dimensional brane embedded isometrically in an D-dimensional bulk space, assuming the matter
fields are confined to the brane. Applying this formalism to cosmology, we derive the generalized Friedmann equations. We derive the density parameter of dark
energy in terms of width of the brane, normal curvature radii and the number
of extra large dimensions. We show that  dark energy could actually
be the manifestation of the local extrinsic shape of the brane. It is shown that
the predictions of this model are in good agreement with observation if we
consider an 11-dimensional bulk space.
\end{abstract}
\section{Introduction}
One of the most sensational discoveries of the past decade is that the expansion
of the Universe is speeding up. This observation finds support in the luminosity
measurements of high redshift supernovae \cite{1} and measurements of degree-scale
anisotropies in the CMB \cite{2}. General relativity leads to a deceleration of the expansion for the Universe if filled with cold matter or radiation.
Since the expansion is speeding up, we are faced with two possibilities:
The first is that 70 percent of the energy density of the Universe exists in a form with large negative pressure, called dark energy. The other possibility is that general relativity
breaks down on the cosmological scales and should be modified or replaced with
a more complicated theory.
Various forms of modification of gravity have been proposed to explain the
late time acceleration of the Universe. In an interesting development, Deffayet,
Dvali and Gabadadze \cite{3} demonstrated that the presence of the 4D Ricci scalar term in the action functional for the brane can lead to a late time
acceleration \cite{4}.  The possibility that branes can naturally produce
a solution to unsolved problems in gravity and cosmology, such
as   dark energy  will generate a great boost
aside from being a significant achievement in the theory and a good verification of the extra dimensions idea.
With all the success that brane gravity models have achieved major
weak points exist as follows;
\begin{itemize}
\item
The nature of phenomenological model building  most of these theories have.
\end{itemize}
\begin{itemize}
\item
Most of  the developments are specific to particular models. For
example, this has given the wrong impression that the brane program is necessarily
a 5D theory based on the AdS5 or Ricci flat bulk.
\end{itemize}
\begin{itemize}
\item
The geometric mechanism, like Kaluza-Klein (KK) and multidimensional gravitational
models, for unification of fundamental forces, using group of isometries
of non-compact bulk space is not well developed.\\
The general theory of relativity provides a geometric description of gravitational
fields. After Einstein work, physicists have been looking for a classical
unified field theory (UFT) which would describe other fundamental interactions in terms
of geometric concepts of spacetime. The first attempt to set up such a UFT
using the concept of adding spacelike extra dimension
was that of Kaluza and Klein \cite{Kaluza}. In KK and its multidimensional
extensions the extra dimensions form a compact coset space $G/H$ for a given
gauge group $G$ and a maximal subgroup $H$. Consequently coordinate transformations
associated with the compact manifold can be interpreted as gauge transformations.\\
In brane scenarios the restriction on the size of the extra dimension is weaker.
Standard model and matter fields are confined to  some hypersurface embedded
in the bulk space. To achieve geometrical unification it is enough to consider
an $n$-dimensional brane with $(n-4)$ compact coset space embedded in $(n+1)$-dimensional bulk space
\cite{b}.  The basic difficulties of such brane models, like the original
KK multidimensional theories, are the compactification of internal space, the problem of vacuum stability for the ground state, the large fermion masses which appear in nonzero modes \cite{Fermion} and the gauge fields appearing
as the dimensional reduction are  not confined to the brane and can propagate in the bulk \cite{Neronov}. To avoid  some of the problems mentioned we need
to find another geometrical origin for the gauge group of Maxwell-Yang-Mills
interactions.
\end{itemize}
\begin{itemize}
\item
A simplifying assumption has been to consider the brane as highly
thin. This approximation is assumed to be valid as long as the energy scales are much smaller
than the energy scales related to the inverse thickness of the brane, or at distance scales much bigger than the thickness of the brane \cite{thin}.\\
Most of the studies are restricted to thin brane models where the standard
model matter fields are infinitely localized on a brane with delta-like distribution.
This kind of models can, however, be only treated as an approximation since
the most fundamental underlying theory (like string and quantum gravity theories)
would have a minimum length scale. Also as we know, for confined standard
model matter fields we have ``hadron barrier'' which could be encountered.
Such barrier could arise because all hadrons couple via the strong interaction
to low mass mesons \cite{Yokawa}. Hence the minimum size of any physical
system containing nucleons cannot be appreciably less than the range of the
strong interactions \cite{Falla}. The final point to note is that in thin
brane approximation, in approaching the brane usually self-interactions of
gravitons are divergent and consequently the justification of confined matter
interactions is impossible. Consideration of a brane thickness in such models
is an essential ingredient to play the role of an effective UV cut-off \cite{Vladimir}.

\end{itemize}
\begin{itemize}
\item
The main difficulty in applying a junction condition is that it is
not unique. Other forms of junction conditions exist, so that
different conditions may lead to different physical results \cite{5}.
Furthermore, these conditions cannot be used in the presence of  more than one
noncompact extra dimension.
\end{itemize}
\begin{itemize}
\item
The Dvali-Gabadadze-Porrati (DGP) brane model which contains self acceleration, the joint constraints from SNLS, BAO, and CMB data shows that this model is disfavored observationally \cite{Sahni}. Moreover the DGP model contains a ghost mode for the branch of the self acceleration \cite{Nicolis}.
\end{itemize}

 In refs.  \cite{Monte} and \cite{6}, the authors proposed a cosmological model for a $4D$ brane embedded in a bulk space with arbitrary large extra
dimensions where the matter fields are confined to the brane. As a result, an extra term in the Friedmann equation appeared that may be interpreted as the X-matter, providing a possible phenomenological explanation for the acceleration of the Universe.  In this paper, we have studied the covariant (model independent) induced Einstein-Yang-Mills field equations on a 4-dimensional brane embedded isometrically in  D-dimensional bulk space. The  embedding is a necessity to make the Riemann curvature more consistent with the purpose  of distinguishing a curved space-time from the flat space of Special Relativity. Indeed, Riemann mentions in his historical paper that his curvature cannot distinguish between a plane or a cylinder or a cone. This problem (which is in fact a topological problem in the sense of shape) was left open and a solution of it was presented by Schl\"afli in 1871, based on the reasoning  that the local curvature cannot be an absolute concept as Riemann proposed. The curvature as a notion of shape is a relative concept, which is  valid only when we have a second object to compare. Therefore, Schl\"afli suggested  the isometric embedding of a manifold into another, so that we could always  compare the curvature of that manifold with the curvature of the embedding space. The isometric condition is to guarantee that the metric of the embedded  manifold is induced by the background. In our model,
the extrinsic curvature plays the role of an independent field in describing dark energy. In fact, to have a common origin for the gravitational
and Yang-Mills interactions we need a $11D$ bulk space and consequently, the Israel junction conditions are not applicable. Also, to include the thickness of the brane we use the Nash's theorem for perturbations of the submanifolds
\cite{7}, where the extrinsic curvature and twist vector fields also play  a dynamical role.
Nash showed that any Riemannian geometry can be generated by a continuous
sequence of infinitesimal deformations defined by the extrinsic curvature.  Some popular brane-world models
use M-theory motivations and use additional postulates such as a $Z_2$ symmetry
across the brane as in the Randall-Sundrum models \cite{8}. This symmetry
was not considered in this paper as essential since the $Z_2$ symmetry breaks
the regularity of the embedding, thus preventing the use of the perturbation
mechanism of Nash's theorem.
\section{The geometrical aspects of model}
\subsection{Locally and isometrically embedding of spacetime}
An effective Einstein-Hilbert action functional for the $4D$ spacetime $({\cal
M}_4,{g})$ embedded in a $D$-dimensional ambient (bulk) space $({\cal M}_D,{\cal G})$ may be derived from the action
\begin{eqnarray}\label{1-1}
-\frac{1}{2\kappa_{D}^2}\int \sqrt{|{\cal G}|}{\cal R}d^Dx-\int{\cal L}_m^*\sqrt{|{\cal
G}|}d^Dx,
\end{eqnarray}
where $\kappa^2_D$ is the bulk space energy scale and ${\cal L}^*_m$ is the
Lagrangian of matter fields confined to the brane with thickness $l$. This
means that we no longer treat the confined energy-momentum tensor as a singular
source, but rather as some smooth distribution along thickness of brane \cite{Carter}.
Variation of action
(\ref{1-1}) respect to the bulk metric ${\cal G}_{AB}$ $(A, B=0, 1, ...,
D-1)$ with signature $(p(-),q(+)),$ we obtain Einstein field equations on the bulk \footnote{Greek indices run from 0 to 3, small case Latin indices run from 4 to $D-1$ and large Latin indices run from 0 to $D-1$. Units so
that $\hbar=c=1$ are used throughout this work.}
\begin{eqnarray}\label{1-2}
G_{AB}=8\pi G^*{\mathcal T}_{AB},
\end{eqnarray}
where $G^*$ is the bulk gravitational constant and ${\cal T}_{AB}$ is the
confined
matter energy-momentum tensor. The confinement hypothesis states that standard
model matter fields are trapped to the $4D$ brane with thickness $l$.
On the other hand, TeV gravitons can  propagate on the bulk space freely.
To construct geometrical properties of the model, we start with the local isometric
embedding structure, consistent with differential structure of Einstein field
equations (\ref{1-2}).

{ To obtain a local and isometric embedding, consider the $D$-dimensional
Lorentzian bulk space $(\mathcal M_D,\mathcal G)$ with local coordinates $\mathcal Y^A=\{\mathcal Y^0,...,\mathcal Y^{D-1}\}$, endowed with a metric
$\mathcal G$. Further, consider in $\mathcal M_D$ a $4D$ submanifold
$(\mathcal M_4,\bar g)$ with local coordinates $x^\mu=\{x^0,...,x^3\}$ and
induced metric $\bar g$. We can then construct an adopted coordinate system
in the bulk space which includes the
local coordinates of submanifold $\mathcal M_4$ as $\{x^\mu,x^a\}$, where $x^a=\{x^{4},...,x^{D-1}\}$
are extrinsic or extra coordinates. In such a condition the submanifold $\mathcal M_4$ is
defined by $x^a=0$.}
Therefore, the isometric local embedding of a given spacetime ${\cal M}_4$ with
induced metric $\bar g_{\mu\nu}$ $(\mu,\nu=0,...,3)$ in an arbitrary bulk
$({\cal M}_D,{\cal G})$, is given by $D$ differential maps
\begin{eqnarray}
{\cal Y}^A: {\cal M}_4\rightarrow {\cal M}_D.
\end{eqnarray}
Also, the vectors
\begin{eqnarray}\label{1-4}
\begin{array}{cc}
\bar{e}_\mu:={\cal Y}^A_{,\mu}\partial_A,\\
\\
\bar{e}_a:={\cal N}^A_a\partial_A,\\
\end{array}
\end{eqnarray}
form a basis of tangent and normal vector spaces respectively at each point
of ${\cal M}_4$, where small Latin indices denotes extra dimensions and ${\cal N}^A_a$ denotes the components of $n=D-4$ unit vector fields orthogonal to the ${\cal M}_4$ and { also normal to each other in direction of the extra coordinates
$x^a$.}
Consequently, differential map ${\cal Y}^A$ satisfies embedding equations
\begin{eqnarray}\label{1-5}
\begin{array}{cc}
{\cal G}(\bar{e}_\mu,\bar{e}_\nu)={\cal G}_{AB}{\cal Y}^A_{,\mu}{\cal Y}^B_{,\nu}=\bar{g}_{\mu\nu},\\
\\
{\cal G}(\bar{e}_\mu,\bar{e}_a)={\cal G}_{AB}{\cal Y}^A_{,\mu}{\cal N}^B_a=0,\\
\\
{\cal G}(\bar{e}_a,\bar{e}_b)={\cal G}_{AB}{\cal N}^A_a{\cal N}^B_b=\eta_{ab},
\end{array}
\end{eqnarray}
where $\eta_{ab}=\epsilon_a\delta_{ab}$, $\epsilon_a=\pm1$ correspond to
two possible signature of each extra dimension \cite{Eisenhart}. Note that these equations
define the normal vector fields $\bar{e}_a$ only up to  $SO(p-1,q-3)$.
If we define the brane projections of covariant derivative of bulk space
with $D_\alpha:={\cal Y}^A_{,\alpha} D_A$, where $D_A$ is the covariant derivative
compatible with ${\cal G}_{AB}$, then the projected gradients of bases vectors
can be decomposed with respect to the bases vectors $\{\bar{e}_\mu,\bar{e}_a\}$ which are the generalizations of well-known Gauss-Weingarten equations
\begin{eqnarray}\label{1-6}
\begin{array}{cc}
D_\mu \bar{e}_\nu=\bar{\Gamma}^\alpha_{\mu\nu}\bar{e}_\alpha+\bar{K}_{\mu\nu}^{\,\,\,\,\,\,\,a}\bar{e}_a,\\
\\
D_\mu \bar{e}_a=-\bar{K}_{\mu\nu a}\bar{e}^\nu+A_{\mu a}^{\,\,\,\,\,\,\,b}\bar{e}_b,
\end{array}
\end{eqnarray}
where $\bar{\Gamma}^\alpha_{\mu\nu}$ are the connection coefficients compatible
with induced metric $\bar{g}_{\alpha\beta}$, $\bar{K}_{\alpha\beta a}$ is
the $a^{th}$ extrinsic curvature (second fundamental form) of the brane
\begin{eqnarray}\label{1-7}
\bar{K}_{\alpha\beta a}=-{\cal G}(D_\alpha \bar{e}_\beta,\bar{e}_a)=-{\cal G}_{AB}{\cal
N}^A_aD_\alpha(\mathcal{Y}^B_{,\beta})=\bar{K}_{\beta\alpha a},
\end{eqnarray}
and $A_{\alpha mn}$ is the extrinsic twist potential (third fundamental form) defined by
\begin{eqnarray}\label{1-8}
A_{\alpha ab}={\cal G}(D_\alpha \bar{e}_a,\bar{e}_b)={\cal G}_{AB}D_\alpha({\cal N}^A_a){\cal
N}^B_b=-A_{\alpha ba}.
\end{eqnarray}
In geometric language, the presence of twist potential tilts
the embedded family of submanifolds with respect to the
normal vector ${\mathcal N}^A$. It is easy to show that if the bulk space has certain
Killing vector fields then $A_{\mu ab}$ transform as the component of a gauge vector field under the group of isometries of the bulk \cite{Shahram}. Note
that in our model the gauge potential can only
be present if the dimension of the bulk space is equal to or
greater than six $(n \geq2)$, because according to  (\ref{1-8}) the twist vector fields $A_{\mu ab}$ are
antisymmetric under the exchange of extra coordinate indices
$a$ and $b$.

The Lie algebra generators of $SO(p-1,q-3)$ are
\begin{eqnarray}\label{algebra1}
J^{mn}=\frac{1}{2}\left(x^m\partial^n-x^n\partial^m\right),
\end{eqnarray}
and the killing vector basis of the space generated by extra dimensions are
\begin{eqnarray}\label{algebra2}
K_{c}^{ab}=J^{ab}(x_c)=x^{[a}\delta^{b]}_c.
\end{eqnarray}
A straightforward calculations show that
\begin{eqnarray}\label{algebra3}
[J_{ab},J_{cd}]=C^{mn}_{abcd}J_{mn},
\end{eqnarray}
where the structure constants are
\begin{eqnarray}\label{algebra4}
C_{abcd}^{mn}=2\delta^m_{[b}\eta_{a][c}\delta^n_{d]}.
\end{eqnarray}

The integrability conditions for embedding (\ref{1-5}) are given by Gauss-Codazzi-Ricci
equations
\begin{eqnarray}\label{1-9}
\begin{array}{cc}
{\cal R}_{ABCD}{\cal Y}^A_\mu{\cal Y}^B_\nu{\cal Y}^C_\rho{\cal Y}^D_\sigma=\bar{R}_{\mu\nu\rho\sigma}-\bar{K}_{\mu\rho
a}\bar{K}_{\nu\sigma}^{\,\,\,\,\ a}+\bar{K}_{\mu\sigma
a}\bar{K}_{\nu\rho}^{\,\,\,\,\ a},\\
\\
{\cal R}_{ABCD}{\cal Y}^A_\mu{\cal N}^B_a{\cal Y}^C_\nu{\cal Y}^D_\rho=\bar{K}_{\mu\nu
a;\rho}-\bar{K}_{\mu\rho a;\nu}-A_{\rho ca}\bar{K}_{\mu\nu}^{\,\,\,\,\,\,c}+A_{\nu ca}\bar{K}_{\mu\rho}^{\,\,\,\,\,\,c},\\
\\
{\cal R}_{ABCD}{\cal N}^A_a{\cal N}^B_b{\cal Y}^C_\mu{\cal Y}^D_\nu=F_{ab\mu\nu}-\bar{K}_{\mu\alpha
a}\bar{K}_{\nu\,\,\,\,\,b}^{\,\,\,\alpha}+\bar{K}_{\nu\alpha
a}\bar{K}_{\mu\,\,\,\,\,b}^{\,\,\,\alpha},
\end{array}
\end{eqnarray}
where ${\cal R}_{ABCD}$, $\bar{R}_{\mu\nu\alpha\beta}$ are the Riemann tensors
of the bulk and brane respectively and
\begin{eqnarray}\label{1-10}
F_{ab\mu\nu}:=A_{\mu ab,\nu}-A_{\nu ab,\mu}-A_{\nu a}^{\,\,\,\,\,\,\,c}A_{\mu
cb}+A_{\mu a}^{\,\,\,\,\,\,\,c}A_{\nu cb},
\end{eqnarray}
is the curvature associated with extrinsic twist vector field \cite{twist}.

\subsection{Geometrical structure of thick brane model}
In most of brane models with single extra non compact dimension a simplifying
assumption has been made to consider the brane hypersurface as  extremely thin. Let us now consider a thick brane of thickness $l$. For simplicity we will restrict
ourselves to the case of constant $l$. One way to generate such thick
brane is to deform the background spacetime $({\cal M}_4,\bar{g}_{\mu\nu})$ in such a way that
it remains compatible with confinement. The embedding functions defined in
(\ref{1-5}) are differentiable
and regular as required by local differentiable embedding.
This follows from Nash's embedding theorem which requires that the embedding functions of the deformed submanifold are represented by convergent positive power series \cite{Poancare}.
According to  Nash's theorem  any submanifold can be generated by
a continuous sequence of small perturbations of an arbitrarily given submanifold. Although it was later generalized  to pseudo Riemannian manifolds
in \cite{7}.
{ Nash's perturbative approach to embedding can be also
described by introducing a small perturbation parameter, $\sigma$, along
an arbitrary direction $\eta$, which is parameterized by $\zeta$ \cite{pert}.
 Hence, starting with the perturbation of
the embedding map (\ref{1-1}) along an
arbitrary transverse direction $\eta $, we obtain the coordinates of a point
not necessarily on $\mathcal M_4$
\begin{eqnarray}\label{1-11}
{\cal Z}^A(x^\mu,\eta)={\cal Y}^A(x^\mu)+\sqrt{\sigma}\zeta(\mathcal{L}_\eta\mathcal{Y})^A={\cal Y}^A(x^\mu)+\sqrt{\sigma}\zeta[\eta,\mathcal{Y}]^A,
\end{eqnarray}
where $\mathcal L$ denotes the Lie derivative.
The tangential projection of $\eta$ can always be identified with the action of brane
diffeomorphism which will subsequently be ignored. Also, one can insert additional
simplification by taking the norm of $\eta$ equal to $\pm1$. Hence, we can consider
a general deformation along the normal  vectors  $\eta=\bar e^a$ parameterized by extra dimensions
$x^a$.
 Consequently
perturbation along orthogonal direction $\bar{e}^a$ will be \cite{pert}
\begin{eqnarray}\label{1-12}
\begin{array}{cc}
\mathcal{Z}^A_{,\mu}(x^\alpha,x^a)=\mathcal{Y}^A_{,\mu}(x^\alpha)+\sqrt{\sigma}x^a\mathcal{N}^A_{a,\mu},\\
\\
\mathcal{N}^A_a\mapsto \mathcal{N}^A_a+\sqrt{\sigma}x^b[\mathcal{N}_a,\mathcal{N}_b]^A=\mathcal{N}^A_a,
\end{array}
\end{eqnarray}
where $\sigma$ denotes a small perturbation parameter.} Thence the deformed embedding will be
\begin{eqnarray}\label{1-13}
\begin{array}{cc}
\mathcal{Z}^A_{,\mu}=\mathcal{Y}^A_{,\alpha}\left(\delta^\alpha_\mu-\sqrt{\sigma}x^a\bar{K}_{\mu\,\,\,\,\,a}^{\,\,\,\,\alpha}\right)+\sqrt{\sigma}\mathcal{N}^A_mA_\mu^{\,\,\,m},\\
\\
\mathcal{Z}^A_{,m}=\sqrt{\sigma}\mathcal{N}^A_m,
\end{array}
\end{eqnarray}
where $A_\mu^{\,\,\,\,m}:=x^aA_{\mu a}^{\,\,\,\,\,\,\,m}$.
Consequently, the set of $\{x^\alpha,x^a\}$ define a Gaussian coordinate system for the bulk space in the  vicinity of $\mathcal{M}_4$. { The
line element of bulk space in the vicinity of brane is $ds^2={\mathcal
G}_{AB}d\mathcal Z^Ad\mathcal Z^B$. Now by substituting Eq. (\ref{1-13}) in the expression for the line
element, the metric of bulk space in the Gaussian coordinates will be }
\begin{eqnarray}\label{1-14}
\mathcal{G}_{AB}=\left(\begin{array}{cc}
g_{\mu\nu}+\sigma A_{\mu c}A_{\nu }^{\,\,\,\,\,c}&  \sigma A_{\mu j} \\
\sigma A_{\nu i} &  \sigma\eta_{ij}
\end{array}\right),
\end{eqnarray}
where
\begin{eqnarray}\label{1-15}
\begin{array}{cc}
g_{\mu\nu}=\bar{g}_{\mu\nu}-2\sqrt{\sigma}x^k\bar{K}_{\mu\nu k}+\sigma
x^mx^n\bar{K}_{\mu\alpha m}\bar{K}_{\nu\,\,\,\,\,n}^{\,\,\,\,\alpha}\\
\\
=\bar{g}^{\alpha\beta}\left(\bar{g}_{\mu\alpha}-\sqrt{\sigma}x^a\bar{K}_{\mu\alpha
a}\right)\left(\bar{g}_{\nu\beta}-\sqrt{\sigma}x^b\bar{K}_{\nu\beta
b}\right),
\end{array}
\end{eqnarray}
represents the metric of the brane with thickness $l$.
As one can see, equation (\ref{1-14}) has the similar appearance to the KK metric where $A_{\mu i}$ plays the role of the Yang-Mills potentials \cite{Yang}. In fact the embedding paradigm suggests another similarity; the extrinsic shape of the brane in the neighbourhood of any of its points is determined with normal curvature \cite{Eisenhart}
\begin{eqnarray}\label{1-16}
 \frac{1}{R_n}=\frac{\bar{K}_{\mu\nu n}\delta x^\mu \delta x^\nu}{\bar{g}_{\mu\nu}\delta
 x^\mu \delta x^\nu}.
 \end{eqnarray}
 The extreme values of normal curvature are calculated by the homogeneous equations
\begin{eqnarray}\label{1-17}
\left(\bar{g}_{\mu\alpha}-l^a\bar{K}_{\mu\alpha
a}\right)\delta x^\mu=0,
\end{eqnarray}
where $l^a_{(\mu)}$ are the curvature radii of the original brane $\bar{g}_{\mu\nu}$
for each principal direction $\delta x^\mu$ and for each normal $e_a$ \cite{Erwin}. Equation
(\ref{1-17}) admits a non trivial solution for curvature directions $\delta
x^\mu$
when
\begin{eqnarray}\label{1-18}
det\left(\bar{g}_{\mu\alpha}-l^a\bar{K}_{\mu\alpha
a}\right)=0.
\end{eqnarray}
It follows that the metric of deformed brane (\ref{1-15}) becomes singular
at the solution of (\ref{1-17}). Consequently according to the (\ref{1-14})
the metric of the bulk space also become singular at the points determined by
this solutions.
Therefore, at each point of $\mathcal{M}_4$ there is a bounded coordinate
space in the normal direction generated with smallest value of the curvature radii. Hence according to the Einstein tube construction, we may associate to each point of brane a closed space ${\mathcal B}_{n}$ generated by the normal curvature
radii.
Consequently, even though the bulk space is not compact, the local shape of the brane generates a local structure line $\mathcal{M}_4\times {\mathcal
B}_{n}$.\\
{ Now in order to reduce the action stated in Eq. (1) in the brane, various
geometrical properties of the bulk space in the vicinity of the brane needs
to be obtained; this procedure is as follows.}
Equation (\ref{1-13}) leads to define the reference frames of the perturbed
geometry as
\begin{eqnarray}\label{1-19}
\begin{array}{cc}
h^A_\mu=\mathcal{Y}^A_{,\alpha}\left(\delta^\alpha_\mu-\sqrt{\sigma}x^a\bar{K}_{\mu\,\,\,\,\,a}^{\,\,\,\,\alpha}
\right),\\
\\
h^A_a=\sqrt{\sigma}\mathcal{N}^A_a,
\end{array}
\end{eqnarray}
and the  tangent and normal vectors to the perturbed brane respectively
by
\begin{eqnarray}\label{1-20}
\begin{array}{cc}
e_\alpha:=\partial_\alpha-A^m_\alpha\partial_m=h^A_\alpha\partial_A,\hspace{.5cm}e^\alpha:=dx^\alpha,\\
\\
e_m:=\frac{1}{\sqrt{\sigma}}\partial_m={\mathcal N}^A_m\partial_A,\hspace{.5cm}e^m:=\sqrt{\sigma}\left(dx^m+A^m_\alpha
dx^\alpha\right).
\end{array}
\end{eqnarray}
Hence, the embedding equations of the perturbed brane will be
\begin{eqnarray}\label{1-21}
\begin{array}{cc}
\mathcal{G}(e_\alpha,e_\beta)=\mathcal{G}_{AB}h^A_\alpha h^B_\beta=g_{\alpha\beta},\\
\\
\mathcal{G}(e_\alpha,e_m)=\mathcal{G}_{AB}h^A_\alpha\mathcal{N}^B_m=0,\\
\\
\mathcal{G}(e_m,e_n)=\mathcal{G}_{AB}\mathcal{N}^A_m\mathcal{N}^B_n=\eta_{mn}.
\end{array}
\end{eqnarray}
Then the projected gradients of bases vectors
can be decomposed with respect to the bases vectors $\{e_\mu,e_a\}$ which are the generalizations of Gauss-Weingarten equations for the perturbed brane

\begin{eqnarray}\label{1-22}
\begin{array}{cc}
D_\mu e_\nu={\Gamma}^\alpha_{\mu\nu}e_\alpha+{K}_{\mu\nu}^{\,\,\,\,\,\,\,a}e_a,\\
\\
D_\mu e_a=-{K}_{\mu\nu a}e^\nu+A_{\mu a}^{\,\,\,\,\,\,\,b}e_b,
\end{array}
\end{eqnarray}
where ${\Gamma}^\alpha_{\mu\nu}$ are the connection coefficients compatible
with induced metric ${g}_{\alpha\beta}$, ${K}_{\alpha\beta a}$ is
the $a^{th}$ extrinsic curvature of the perturbed brane and $D_\alpha:=h^A_\alpha
D_A$.
Now using the Koszul formula
\begin{eqnarray}\label{1-23}
\begin{array}{cc}
2\mathcal{G}(D_ZY,X)=-X\mathcal{G}(Y,Z)+Y\mathcal{G}(Z,X)+Z\mathcal{G}(X,Y)-\mathcal{G}([Z,X],Y)\\
\\
-\mathcal{G}([Y,Z],X)+\mathcal{G}([X,Y],Z);\hspace{.4cm} X,Y,Z\epsilon{\cal
X}_p({\cal M_D}),
\end{array}
\end{eqnarray}
the Gauss-Weingarten equations (\ref{1-22}) and the fact that the commutator of two tangent vectors no longer vanishes
\begin{eqnarray}\label{1-24}
[e_\alpha,e_\beta]=\sqrt{\sigma}x^aF^n_{\,\,\,a\alpha\beta}e_n,
\end{eqnarray}
one can easily obtain the components of connection and extrinsic curvature
\begin{eqnarray}\label{1-25}
\begin{array}{cc}
\Gamma^\rho_{\alpha\beta}=\frac{1}{2}g^{\rho\gamma}\left(\hat{\partial}_\alpha
g_{\gamma\beta}+\hat{\partial}_\beta g_{\gamma\alpha}-\hat{\partial}_\gamma
g_{\alpha\beta}\right)=\\
\\
{\bar{\Gamma}^\rho_{\alpha\beta}}-\sqrt{\sigma}x^m\bar{g}^{\rho\mu}\left(\nabla_\alpha^{(tot)}\bar{K}_{\mu\beta
m}+\nabla_\beta^{(tot)}\bar{K}_{\mu\alpha m}-\nabla_\mu^{(tot)}\bar{K}_{\alpha\beta
m}\right)\\
\\:={\bar{\Gamma}^\rho_{\alpha\beta}}-\sqrt{\sigma}x^mH^\rho_{\alpha\beta
m},
\end{array}
\end{eqnarray}
and
\begin{eqnarray}\label{1-26}
\begin{array}{cc}
K_{\alpha\beta a}=-\frac{1}{2\sqrt{\sigma}}\partial_ag_{\alpha\beta}+\frac{\sqrt{\sigma}}{2}x^bF_{ab\alpha\beta}\\
\\
=\bar{K}_{\alpha\beta a}-\frac{\sqrt{\sigma}}{2}x^m\left(\bar{K}_{\alpha\gamma m}\bar{K}^\gamma_{\,\,\,\beta
a}+\bar{K}_{\alpha\gamma a}\bar{K}^\gamma_{\,\,\,\beta
m}-F_{\alpha\beta am}\right),
\end{array}
\end{eqnarray}
where
\begin{eqnarray}\label{1-27}
\begin{array}{cc}
\nabla_\mu^{(tot)}\bar{K}_{\alpha\beta m}:=\bar{K}_{\alpha\beta m;\mu}-A_{\mu
mn}\bar{K}_{\alpha\beta}^{\,\,\,\,\,\,n},\\
\\
\hat{\partial}_\alpha:=\partial_\alpha-A^m_\alpha\partial_m,
\end{array}
\end{eqnarray}
denote the total covariant  (gauge covariant) derivative and horizontal lift
of the base vector fields respectively.
Now the geometric meaning of the second fundamental form (\ref{1-26}) is quite clear: its symmetric part shows that the second fundamental form propagates in the bulk. The antisymmetric part is proportional to a Yang-Mills
gauge field, which really can be thought of as a kind of curvature.
The integrability conditions for the deformed brane are the Gauss, Codazzi
and Ricci equations, respectively
\begin{eqnarray}\label{1-28a}
\langle e_\mu,{\cal R}(e_\alpha,e_\beta)e_\gamma\rangle=R_{\mu\gamma\alpha\beta}+K_{\alpha\gamma}^{\,\,\,\,\,\,\,a}K_{\beta\mu
a}-K_{\beta\gamma}^{\,\,\,\,\,\,a}K_{\alpha\mu a},
\end{eqnarray}
\begin{eqnarray}\label{1-28b}
\langle e_l,{\cal R}(e_\alpha,e_\beta)e_\gamma\rangle=
\hat\nabla_\alpha K_{\beta\gamma l}-\hat\nabla_\beta K_{\alpha\gamma l}-K_{\alpha\gamma}^{\,\,\,\,\,\,\,\,a}A_{\beta
al}+K_{\beta\gamma}^{\,\,\,\,\,\,\,\,a}A_{\alpha al},
\end{eqnarray}
\begin{eqnarray}\label{1-28c}
\langle e_l,{\cal R}(e_\alpha,e_\beta)e_a\rangle
=K_{\alpha\,\,\,\,a}^{\,\,\,\,\nu}K_{\beta al}-K_{\beta\,\,\,\,a}^{\,\,\,\,\nu}K_{\alpha
al}-F_{\beta\alpha al,}
\end{eqnarray}
where $\hat\nabla$ denotes the covariant derivative defined  as
\begin{eqnarray}\label{1-29}
\hat\nabla_\mu T^{...}_{...\nu...}=\hat\partial_\mu T^{...}_{...\nu...}+...-\hat\Gamma_{\mu\nu}^\alpha
T^{...}_{...\alpha...},\hspace{.5cm}\hat\nabla{g}_{\alpha\beta}=0,
\end{eqnarray}
and $R_{\alpha\beta\gamma\mu}$ is the Riemann tensor of the deformed brane
\begin{eqnarray}\label{1-30}
\begin{array}{cc}
R^\sigma_{\,\,\,\gamma\alpha\beta}=\hat\partial_\alpha\Gamma^\sigma_{\beta\gamma}-\hat\partial_\beta\Gamma^\sigma_{\alpha\gamma}
+\Gamma^\rho_{\beta\gamma}\Gamma^\sigma_{\alpha\rho}-\Gamma^\rho_{\alpha\gamma}\Gamma^\sigma_{\beta\rho}+
\sqrt{\sigma}x^aF^n_{\,\,\,\,a\alpha\beta}K_{\gamma\,\,\,\,\,n}^{\,\,\,\,\sigma}\\
\\
=\bar{R}^\sigma_{\gamma\alpha\beta}+\sqrt{\sigma}x^m\left(\nabla^{(tot)}_\beta
H^\sigma_{\alpha\gamma m}-\nabla^{(tot)}_\alpha
H^\sigma_{\beta\gamma m}+\bar{K}_{\gamma\,\,\,\,n}^{\,\,\,\sigma}F^n_{\,\,\,\,m\alpha\beta}\right)\\
\\
-\frac{\sigma}{2}x^mx^nF_{\gamma\,\,\,\,\,am}^{\,\,\,\sigma}F_{\alpha\beta\,\,\,\,n}^{\,\,\,\,\,\,\,a}+{\cal
O}(\sigma l^2/L^2).
\end{array}
\end{eqnarray}
It is worth noticing that the Riemann tensor of the deformed brane is a reducible tensor.
Again using Koszul formula we obtain
\begin{eqnarray}\label{1-30a}
\begin{array}{cc}
h^A_aD_Ae_\mu:=D_ae_\mu=-K_{\mu\,\,\,\,a}^{\,\,\,\,\nu}e_\nu,\\
\\
D_ae_b=0,
\end{array}
\end{eqnarray}
which helps us to obtain the remaining group of higher dimensional Riemann tensor
components
\begin{eqnarray}\label{1-30b}
\mathcal{R}_{DCAB}h^D_\mu h^A_\alpha
\mathcal{N}^B_m\mathcal{N}^C_n=g_{\mu\sigma}\left[\frac{1}{\sqrt{\sigma}}K_{\alpha\,\,\,\,\,n,m}^{\,\,\,\,\sigma}-K_{\alpha\,\,\,\,\,n}^{\,\,\,\,\nu}K_{\nu\,\,\,\,\,m}^{\,\,\,\,\sigma}\right],
\end{eqnarray}
\begin{eqnarray}\label{1-30bb}
\mathcal{R}_{DCAB}\mathcal{N}^D_lh^A_\alpha\mathcal{N}^B_m\mathcal{N}^C_n=0,
\end{eqnarray}
\begin{eqnarray}\label{1-30bbb}
\mathcal{R}_{DCAB}\mathcal{N}^A_ah^B_\beta
h^C_\gamma h^D_\mu=g_{\mu\sigma}\left[\frac{1}{\sqrt{\sigma}}\Gamma^\sigma_{\beta\gamma,a}+\hat\nabla_\beta
K_{\gamma\,\,\,\,a}^{\,\,\,\,\sigma}-A_{\beta a}^{\,\,\,\,\,\,\,b}K_{\gamma\,\,\,\,b}^{\,\,\,\,\sigma}\right],
\end{eqnarray}
\begin{eqnarray}\label{1-30bc}
\mathcal{R}_{DCAB}\mathcal{N}^D_lh^C_\gamma\mathcal{N}^A_ah^B_\beta=\frac{1}{\sqrt{\sigma}}K_{\beta\gamma l,a}+K_{\gamma\,\,\,\,a}^{\,\,\,\,\nu}K_{\beta\nu
l},
\end{eqnarray}
\begin{eqnarray}\label{1-30bd}
\mathcal{R}_{DCAB}\mathcal{N}^A_nh^B_\beta
\mathcal{N}^C_ah^D_\mu=g_{\alpha\mu}\left[-\frac{1}{\sqrt{\sigma}}K_{\beta\,\,\,\,\,a,n}^{\,\,\,\,\alpha}+K_{\beta\,\,\,\,\,a}^{\,\,\,\,\nu}K_{\nu\,\,\,\,\,n}^{\,\,\,\,\alpha}\right].
\end{eqnarray}
Hence, the reduction of the Riemann tensor of the bulk space provides a
 generalization of
 the Gauss, Codazzi and Ricci equations and remaining components.
Contraction of the Gauss equation  (\ref{1-28a}), using equations (\ref{1-30b})
and(\ref{1-30bc}) gives the Ricci scalar
\begin{eqnarray}\label{1-31}
\begin{array}{cc}
\mathcal{R}=R+K_{\alpha\beta a}K^{\alpha\beta a}-K^aK_a+\frac{1}
{\sqrt{\sigma}}g^{\mu\nu}K^{\,\,\,\,\,\,m}_{\nu\mu\,\,\,\,\,,m}+\frac{1}{\sqrt{\sigma}
}K^a_{,a}\\
\\
\simeq\bar{R}+\bar{K}_{\alpha\beta a}\bar{K}^{\alpha\beta a}-\bar{K}_a\bar{K}^a+\sqrt{\sigma}x^m(\bar{g}^{\gamma\beta}\nabla^{(tot)}_\beta
H^\sigma_{\sigma\gamma m}\\
\\
-\bar{g}^{\gamma\beta}\nabla^{(tot)}_\sigma H^\sigma_{\beta\gamma
m} +2\bar{K}^{\alpha\beta}_{\,\,\,\,m}\bar{R}_{\alpha\beta})-\frac{\sigma}{4}x^mx^nF^{\alpha\beta}_{\,\,\,\,\,\,am}F_{\alpha\beta\,\,\,\,\,\,n}^{\,\,\,\,\,\,\,\,a}.
\end{array}
\end{eqnarray}
As was mentioned earlier, the physical space on the bulk space is restricted
to the Einstein tube.
\section{Induced gravitational field equations on a thick brane}
 Now if  we assume { a flat bulk space with} all extra dimensions
 being spacelike , then ${\mathcal B}_{n}$ may be taken to be the $n$-sphere
\begin{eqnarray}\label{1-32}
S_{n}=SO(n)/SO(n-1),
\end{eqnarray}
with radius  $L:=min\{l^a_{(\mu)}\}$. Consequently by inserting
\begin{eqnarray}\label{1-33}
\sqrt{|g|}\simeq\sqrt{|\bar g|}\left(1-\sqrt{\sigma}x^m\bar{K}_m+\frac{1}{2}\sigma
x^m x^n\left(\bar{K}_m{\bar K}_n-{\bar K}_{\alpha\beta m}{\bar K}^{\alpha\beta}_{\hspace{.4cm}n}\right)\right),
\end{eqnarray}
and  equation  (\ref{1-31}) into the action functional (\ref{1-2}) and remembering
the confinement hypothesis (matter and gauge fields are confined to the brane
with thickness $l$ while gravity could  propagate in the bulk space up
to the curvature radii $L$ { with the geometry of $S_n$}) we obtain
\begin{eqnarray}\label{1-34}
\begin{array}{cc}
-\frac{1}{2\kappa^2_D}\int{\sqrt{|\mathcal{G}|}{\cal R}d^Dx}\simeq\\
\\
-\frac{M_D^{n+2}\sigma^{n/2}V_n}{16\pi}\int
L^n\sqrt{|\bar{g}|}\left(\bar{R}-\bar{K}_{\alpha\beta m}{\bar K}^{\alpha\beta m}
+ \bar{K}^2\right)d^4x\\
\\
+\frac{nM_D^{n+2}\sigma^{n/2}l^3V_n}{64\pi(n+2)}\int L^{n-1}\sqrt{|\bar{g}|}F_{\mu\nu
mn}F^{\mu\nu mn}d^4x,
\end{array}
\end{eqnarray}
where $V_n=\pi^{\frac{n}{2}}/\Gamma(\frac{n}{2}+1)$. Hence, the relation between the fundamental scale $M_D$ and the $4D$ Planck
scale $M_{Pl}$ will be
\begin{eqnarray}\label{1-35}
M_{Pl}^2=\frac{\pi^{n/2}}{\Gamma(n/2+1)}L^n{\sigma}^{n/2}M_D^{n+2}.
\end{eqnarray}
On the other hand, with an eye on the Lagrangian density of the Yang-Mills field
\cite{Scheck}
\begin{eqnarray}\label{f1}
{\mathcal L}^{(YM)}=\frac{1}{4g_i^2K}tr(F_{\mu\nu}F^{\mu\nu}),
\end{eqnarray}
the last part of (\ref{1-34}) gives
\begin{eqnarray}\label{1-15b}
\frac{4\pi}{Kg^2_i}=\frac{nV_nM_D^{n+2}\sigma^{n/2}l^3L^{n-1}}{4(n+2)},
\end{eqnarray}
where $g_i$ are the gauge couplings, and the index $i$ labels the simple subgroups of the gauge group. The positive constant $K$ depends on the representation
symmetry group but dos not depend on the generators of Lie algebra. In adjoint
representation, it can be calculated from \cite{Scheck}
\begin{eqnarray}\label{k}
C^{mn}_{abcd}C^{cd}_{a'b'mn}=K^{(ad)}\left[\delta_{ab'}\delta_{a'b}-\delta_{aa'}\delta_{bb'}\right].
\end{eqnarray}
Equations (\ref{algebra4}) and (\ref{k})  immediately give
\begin{eqnarray}\label{k1}
K^{(ad)}=\begin{cases}2,\hspace{1.4cm} n=2,\\
{2}(n-2),\hspace{.2cm}n>2. \\
\end{cases}
\end{eqnarray}
The induced $4D$ gravitational ``constant'' is dependent on the normal curvature
radii and consequently is not constant. This implies  violation of the strong equivalence principle (SEP) on the brane. In scalar-tensor theories
of gravity the gravitational constant having a dimension of
length squared emerges through a cosmological background value of the scalar field. Also in KK and multidimensional gravitational theories with compact extra dimensions, $4D$
gravitational constant is a function of the scale factor of the extra dimensions. While in our approach the extra dimensions are not compact and  emerge through external shape of the brane.
Equations (\ref{1-35}) and (\ref{1-15b}) immediately give us the following fundamental
relation between normal curvature radii,  thickness of the brane, number of
extra dimensions and $4D$ Planck's mass
\begin{eqnarray}\label{1-L}
L=\frac{n}{4(n+2)}l^3M_{Pl}^2K^{(ad)}\frac{g_i^2}{4\pi}.
\end{eqnarray}
Note that if $L\sim l\sim l_{Pl}$, the above equation reduces to the equivalent  relation in KK gravity \cite{Wesson}. To obtain the induced field equations, we only have to write the Einstein's
field equations (\ref{1-2}) in the Gaussian frame $\{h^A_\mu,{\cal N}^A_a\}$
of the bulk space. After some calculations one can obtain three sets of field equations
\begin{eqnarray}\label{1-36}
\begin{array}{cc}
R_{\alpha\beta}-\frac{1}{2}g_{\alpha\beta}R+\left(K^{\mu\,\,\,\,\,a}_{\,\,\,\,\,\alpha}+K_\alpha^{\hspace{.3cm}\mu
a}\right)K_{\beta\mu a}-K_{\beta\alpha}^{\hspace{.4cm}a}K_a+\\
\\
+\frac{1}{\sqrt{\sigma}}\eta^{ab}K_{\beta\alpha a,b}-\frac{1}{2}g_{\alpha\beta}(K^{\mu\nu
a}K_{\mu\nu a}-K^aK_a+\frac{1}{\sqrt{\sigma}}g^{\mu\nu}\eta^{mn}K_{\mu\nu m,n}+\\
\\
+\frac{1}{\sqrt{\sigma}}K^m_{\,\,\,,m})=8\pi G^*\mathcal{T}_{AB}h^A_\alpha h^B_\beta,\\
\\
\hat\nabla_\beta K_a-\hat\nabla_\alpha K_{\beta\hspace{.2cm}a}^{\,\,\,\,\alpha}-A_{\beta
a}^{\hspace{.4cm}b}K_b+A_{\alpha a}^{\hspace{.4cm}b}K_{\beta \hspace{.2cm}b}^{\hspace{.2cm}\alpha}=8\pi
G^*\mathcal{T}_{AB}\mathcal{N}^A_ah^B_\beta,\\
\\
\frac{1}{\sqrt{\sigma}}K_{a,b}-K^{\alpha\beta}_{\hspace{.4cm}a}K_{\beta\alpha
b}-\frac{1}{2}\eta_{ab}(R+K^{\mu\nu m}K_{\mu\nu m}-K_mK^m+\\
\\
+\frac{1}{\sqrt{\sigma}}g^{\mu\nu}g^{mn}K_{\mu\nu m,n}+\frac{1}{\sqrt{\sigma}}g^{mn}K_{m,n})=8\pi
G^*\mathcal{T}_{AB}\mathcal{N}^A_a\mathcal{N}^B_b.
\end{array}
\end{eqnarray}
In most general cases, without assuming the existence of matter fields in the
bulk space, the confinement hypothesis states that
\begin{eqnarray}\label{1-37}
\begin{array}{cc}
T_{\alpha\beta}:=\sigma^{\frac{n}{2}}\int\mathcal{T}_{AB}\mathcal{Y}^A_{,\alpha}\mathcal{Y}^B_{,\beta}d^{n}x,\\
\\
\Xi_{\alpha a}:=\sigma^{\frac{n}{2}}\int\mathcal{T}_{AB}\mathcal{Y}^A_{,\alpha}\mathcal{N}^B_{a}d^nx,\\
\\
\Xi_{ab}:=\sigma^{\frac{n}{2}}\int\mathcal{T}_{AB}\mathcal{N}^A_{a}\mathcal{N}^B_{b}d^nx.
\end{array}
\end{eqnarray}
 The field equations (\ref{1-36})
are equivalent to the Einstein field equations (\ref{1-2}) representing any
regular perturbation in the domain of the brane width. To establish effective
induced gravitational equations for a observer living in the brane, the
$4D$ material (\ref{1-37}) and geometrical quantities are defined as an average
over the thickness of the brane. Hence to obtain the field equations
for the covariant thick brane world model, it is sufficient to integrate field
equations (\ref{1-36}) along extra dimensions. Consequently, using (\ref{1-37})
and (\ref{1-35}), in the first approximation the field equations will be
\begin{eqnarray}\label{1-38a}
\bar{G}_{\alpha\beta}= -Q_{\alpha\beta}+8\pi G_N\left(T_{\alpha\beta}+T_{\alpha\beta}^{(YM)}\right),
\end{eqnarray}
\begin{eqnarray}\label{1-38b}
\nabla^{(tot)}_\beta\bar{K}_a-\nabla^{(tot)}_\alpha\bar{K}_{\beta a}^{\alpha}=8\pi
G_N\Xi_{a\beta} ,
\end{eqnarray}
\begin{eqnarray}\label{1-38c}
\begin{array}{cc}
\frac{G_N}{Kg_i^2}\left(F^{\alpha\beta}_{\,\,\,\,\,\,\,am}F_{\alpha\beta
b}^{\,\,\,\,\,\,\,\,\,\,m}+\frac{1}{2}\eta_{ab}F_{\alpha\beta}^{\hspace{.3cm}lm}F^{\alpha\beta}_{\hspace{.3cm}lm}\right)
-
\\
-\frac{1}{2}\eta_{ab}\left(\bar{R}+\bar{K}_{\mu\nu m}\bar{K}_{\mu\nu}^{\hspace{.3cm}m}-\bar{K}_a\bar{K}^a\right)=8\pi
G_N\Xi_{ab},
\end{array}
\end{eqnarray}
where $\bar{G}_{\alpha\beta}$ is the induced Einstein tensor, $T_{\alpha\beta}^{(YM)}$ is the Yang-Mills energy-momentum tensor
\begin{eqnarray}\label{ym}
T_{\alpha\beta}^{(YM)}=\frac{1}{4\pi Kg_i^2}\left[F_{\alpha\,\,\,\,lm}^{\,\,\,\sigma}F_{\beta\sigma}^{\,\,\,\,\,lm}-\frac{1}{4}\bar{g}_{\alpha\beta}F_{\mu\nu
lm}F^{\mu\nu lm}\right],
\end{eqnarray}
and $Q_{\alpha\beta}$ is a conserved quantity
defined as
\begin{eqnarray}\label{1-39}
\begin{array}{cc}
Q_{\alpha\beta}:=\bar{g}^{\gamma\eta}\eta^{ab}\bar{K}_{\alpha\gamma a}\bar{K}_{\beta\eta b}-\eta^{ab}\bar{K}_a\bar{K}_{\alpha\beta b}-\frac{1}{2}\bar{g}_{\alpha\beta}\left(\bar{K}^{\mu\nu
a}\bar{K}_{\mu\nu a}-\bar{K}_a\bar{K}^a\right),\\
\\
Q^{\alpha\beta}_{\,\,\,\,\,\,\,\,;\beta}=0,
\end{array}
\end{eqnarray}
with direct consequence that the product of $4D$ gravitational ``constant'' and
the induced energy-momentum tensor { for confined matter and Yang-Mills fields are conserved.}
\begin{eqnarray}\label{1-41a}
\left(G_NT^{\alpha\beta}\right)_{;\beta}=\left(G_NT^{\alpha\beta}_{(YM)}\right)_{;\beta}=0.
\end{eqnarray}
{It is worth stating that the gravitational ``constant'' ($G_N$) is obtained from Eq. (\ref{1-35}), where the curvature radius on the LHS of Eq. (\ref{1-35}) is resulted
from Eq. (\ref{1-18}). Hence Eq. (\ref{1-18}) plays an auxiliary role on
the field equations.  The extrinsic curvature can be calculated from Eqs. (\ref{1-38b}) and (\ref{1-39}).
In addition, the twist potential can be obtained from Eq. (\ref{1-41a}).
Equation (\ref{1-38c}) puts a restriction on the twisting vector fields. In fact
by determining the trace of Eq. (\ref{1-38a}) which gives the Ricci scalar, and plugging
it in Eq (\ref{1-38c}), we have
\begin{eqnarray}\label{rest}
\frac{1}{Kg_i^2}\left(F^{\alpha\beta}_{\,\,\,\,\,\,\,am}F_{\alpha\beta
b}^{\,\,\,\,\,\,\,\,\,\,m}+\frac{1}{2}\eta_{ab}F_{\alpha\beta}^{\hspace{.3cm}lm}F^{\alpha\beta}_{\hspace{.3cm}lm}\right)
-8\pi\left(\Xi_{ab}-\frac{1}{2}T\eta_{ab}\right)=0.
\end{eqnarray}
 Note that this is not a differential equation, but is only a restriction on the twisting vector field. This is equivalent to the equations obtained for the KK theory when the scale factor of compact space is constant, see Ref. \cite{Wesson}. To construct an Einstein-Yang-Mills-Scalar theory similar to
the KK theory
(where the LHS of equation (\ref{1-38c}) plays the role of source term) it
is enough to consider a brane model with non-constant thickness. In this
paper we considered a brane model for a weak field approximation with constant
thickness for brane to illustrate  how the extrinsic shape of brane can explain the late time acceleration of the universe.} 

\section{The FLRW brane}
\subsection{Induced field equations on FLRW brane}
Phenomenological fluid models of dark energy are difficult to motivate. Usually
such models are unstable under perturbations, since the adiabatic speed of sound is imaginary for negative $\omega=p/\rho$
\begin{eqnarray}
c_s^2=\frac{\dot p}{\dot\rho}=\omega-\frac{\dot\omega}{3H(1+\omega)}.
\end{eqnarray}
Hence for constant and negative values of $\omega$ the model is physically not viable. In what follows we will construct a geometrical fluid with constant
$\omega$ as a replacement of the phenomenological fluid.

First, we will analyze the influence of the extrinsic curvature
terms on a FLRW Universe  when  all extra dimensions are spacelike. Also for simplicity,
we assume that the twisting vector fields $A_{\mu ab}$ vanish. Recall that the standard FLRW line element is spatially homogeneous and isotropic
which can be written as
\begin{eqnarray}\label{2-1}
ds^2=-dt^2+a(t)^2\left(\frac{dr^2}{1-kr^2}+r^2d\Omega^2\right),
\end{eqnarray}
where $a(t)$ is the cosmic scale factor and $k$ is $+1$ $-1$ or $0$, corresponding
to the closed, open, or flat Universes. To be consistent with symmetries
of the embedded FLRW Universe, the  energy-momentum tensor, ${\mathcal T}_{AB}$,
must be diagonal. Hence, in the simplest realization, we have
\begin{eqnarray}\label{2-2}
\begin{array}{cc}
T_{\mu\nu}= (\rho+p)u_\mu u_\nu+pg_{\mu\nu},\\
\\
\Xi_{\mu a}=0,\\
\\
\Xi_{ab}=p_{ext.}\eta_{ab},
\end{array}
\end{eqnarray}
where $p_{ext.}$ is the pressure along the extra dimensions. The general
solution of equations (\ref{1-38b}) is
\begin{eqnarray}\label{2-3}
\begin{array}{cc}
\bar{K}_{00m}=-\frac{1}{a(t)H}\frac{d}{dt}\left(\frac{f_m(t)}{a(t)}\right),\\
\\
\bar{K}_{\alpha\beta m}=\frac{f_m(t)}{a^2(t)}\bar{g}_{\alpha\beta},\hspace{.5cm}\alpha,\beta=1,2,3,
\end{array}
\end{eqnarray}
where $f_m(t)$ are arbitrary functions of the cosmic time $t$, and $H$ is the
Hubbel parameter. The symmetries of spacetime encourage to assume that the functions
$f_m(t)$ are considered as equal; $f_m(t)=f(t)$. If we set $\phi(t):=a^2(t)/f(t)$ and
$h:=\dot{\phi}(t)/\phi(t)$, the components of $Q_{\alpha\beta}$ defined in
(\ref{1-39}) will be
\begin{eqnarray}\label{2-4}
\begin{array}{cc}
Q_{00}=\frac{3n}{\phi^2},\\
\\
Q_{\alpha\beta}=-\frac{n}{\phi^2}\left(3-\frac{2h}{H}\right)\bar{g}_{\alpha\beta},
\hspace{.3cm}\alpha,\beta=1,2,3.
\end{array}
\end{eqnarray}
where $n:=D-4$. { It can be directly verified that the conservation of $Q_{\alpha\beta}$
is  satisfied trivially without bringing up a new differential equation for
obtaining $\phi$.} To obtain $4D$ gravitational constant, it is necessary to obtain the extrinsic curvature radii of FLRW Universe from equation (\ref{1-18}). One can easily
show
\begin{eqnarray}\label{2-5}
\begin{array}{cc}
l_{(0)}^m=\frac{a^2(t)}{f_m(t)}=\phi(t),\\
\\
l_{(\alpha)}^m=\frac{\phi(t)}{1-\frac{h}{H}}, \hspace{.3cm}\alpha=1,2,3.
\end{array}
\end{eqnarray}
Thence, if we assume $h/H>0$, the normal curvature radii will be
\begin{eqnarray}\label{2-6}
L=min(l_{(0)},l_{(\alpha)})=\phi(t).
\end{eqnarray}
Therefore according to  (\ref{1-35}) the induced gravitational ``constant'' is not actually a true constant
and  dependents on the local normal radii of the FLRW submanifold as
\begin{eqnarray}\label{2-7}
G_N=G_0\left(\frac{\phi(t)}{\phi_0}\right)^{-n},
\end{eqnarray}
where $\phi_0$ is the present value of $\phi(t)$ and
\begin{eqnarray}\label{2-8}
G_0:=V_n^{-1}\phi_0^{-n}\sigma^{-n/2}M_D^{-(n+2)},
\end{eqnarray}
is the  $4D$ gravitational constant at the present epoch. Note that the behaviour
of $L$ near a spacetime singularity depends on the topological nature of
that singularity. In the point-like singularity (like FLRW case) all values
of $l^\mu_{(a)}$ tend to zero so that $L$ also tends to zero.
Therefore by inserting (\ref{2-4}) and (\ref{2-7}) into  (\ref{1-38a}),
the Friedmann equations will be
\begin{eqnarray}\label{2-9}
\begin{array}{cc}
H^2+\frac{k}{a^2}=\frac{8\pi G_0}{3}\left(\frac{\phi_0}{\phi}\right)^n\rho+\frac{n}{\phi^2},\\
\\
\frac{\ddot{a}}{a}=-\frac{4\pi G_0}{3}\left(\frac{\phi_0}{\phi}\right)^n(\rho+3p)+\frac{n}{\phi^2}\left(1-\frac{h}{H}\right).\\
\end{array}
\end{eqnarray}
Also Eq. (\ref{rest}) gives
\begin{eqnarray}\label{2-9b}
3p-\rho-2p_{ext.}=0,
\end{eqnarray}
{which shows that the pressure along the thickness of brane is proportional
to the pressure and energy density of ordinary matter. In fact, Eq. (\ref{2-9b}) is due to the constant thickness of the brane. indicates the is an affect of constant thickness of brane. In general the RHS of this equation
is proportional to the dynamics of thickness, and consequently the LHS
is a source for the dynamics of the thickness. Our assumption on the constancy of
thickness hence gives a simple restriction on the pressure component of
the confined matter fields along the extra dimensions.}
 Also the modified conservation  of the  energy-momentum
tensor (\ref{1-41a}) plus equation of state for prefect fluid $p=\omega\rho$ implies
\begin{eqnarray}\label{2-10a}
G_N\rho=G_0\rho_0\left(\frac{a(t)}{a_0}\right)^{-3(1+\omega)},
\end{eqnarray}
{where by inserting (\ref{2-7}) on it gives}
\begin{eqnarray}\label{bbb111}
\rho=\rho_0\left(\frac{a}{a_0}\right)^{-3(1+\omega)}\left(\frac{\phi(t)}{\phi_0}\right)^n.
\end{eqnarray}
The Friedmann equations (\ref{2-9}) depends  on the radial
bending function $\phi(t)$ which remains arbitrary and dose not have a corresponding
differential equation.
The field equations in (\ref{1-38b}) are first order partial differential equations and
 do not determine uniquely the components of the extrinsic curvature.
According to  (\ref{1-7}), the extrinsic curvature is dependent
on the metrical structure and embedding functions $\mathcal{Y}^A(x^\alpha)$.
Therefore, the components of the extrinsic curvature can be calculated uniquely
when the metrics of bulk and  brane spaces are uniquely  specified. {
In brane models, we have only one non-compact extra dimension, where by using the junction conditions one can obtain the components of extrinsic curvature. But here we have more than one extra dimension and consequently it is impossible
 to obtain the extrinsic curvature with the same methods. Hence to obtain
 the components of extrinsic curvature we need extra assumptions or conditions.
 For example if we assume that the SEP is confirmed on the brane, the
 gravitational constant would be really  a constant and therefore the curvature radii
 will be fixed. This makes the bending function $\phi$ to be constant. Note
 that it has been also stated that the extrinsic curvature may satisfies the external Gupta's equations \cite{Cap} as extra condition.
  }In Friedmann equations (\ref{2-9}), practically we need the curvature radii $L=\phi(t)$ not the extrinsic curvature components.

According to  relation (\ref{2-7}) the gravitational constant is a function
of cosmic time $t$ and consequently the SEP is violated
on the brane. The SEP comprises two assumption:  that the weak equivalence
principle holds, and that all local  experiments (gravitational or
non-gravitational) are independent of where and when in the Universe they
are performed. If SEP holds on the FLRW brane, then $G_N$ must be constant
or equivalently the brane will be umbilic.  Furthermore, one experiment is not sufficient
to test the SEP; at last two experiments are needed at two different times.
Hence what is physically relevant is $\dot{G}_N/G_N$.

Using equation (\ref{2-7}) we have
\begin{eqnarray}\label{2-11}
\frac{\dot{G}_N}{G_N}=-nh.
\end{eqnarray}
Several constraints on the local rate of change of $G_N$ using
 observation of lunar occultations and eclipses, gravitational lensing,   evolution of the Sun,  planetary and lunar radar-ranging
measurements, Viking landers or data from the binary pulsar PSR 1913+16 have been obtained up to now. Among these measurements, the last one provided the most reliable upper bound \cite{Damour}
\begin{eqnarray}\label{s1}
-(1.10\pm1.07)\times10^{-11}\mathrm{yr}^{-1} <\frac{\dot{G_N}}{G_N}< 0.
\end{eqnarray}
On the other hand, the best upper bound has been obtained using helioseismological data \cite{Ku}
\begin{eqnarray}\label{s2}
-1.6\times10^{-12}\mathrm{yr}^{-1} <\frac{\dot{G}_N}{G_N}< 0.
\end{eqnarray}
At cosmological distances the best upper
bound comes from the Hubble diagram of distant type Ia supernovae \cite{Gazta}
\begin{eqnarray}\label{s3}
-10^{-11}\mathrm{yr}^{-1}\leq\frac{\dot{G}_N}{G_N}<0, \hspace{.5cm}\mathrm{z}\approx0.5.
\end{eqnarray}
All these upper bounds are negative and shows that $G_N$  in the simplest model may  vary as \cite{Hellings}
\begin{eqnarray}\label{2-12}
\frac{\dot{G}_N}{G_N}=-\delta H,
\end{eqnarray}
where $\delta$ is a small positive dimensionless constant. This assumption with (\ref{2-11}) leads
us to
\begin{eqnarray}\label{2-13}
L=\phi(t)=\phi_0\left(\frac{a(t)}{a_0}\right)^{\frac{\delta}{n}}.
\end{eqnarray}
Inserting (\ref{2-13}) into  (\ref{2-9}), the Freidmann equations will
be
\begin{eqnarray}\label{2-14}
\begin{array}{cc}
H^2+\frac{k}{a^2}=\frac{8\pi G_0}{3}\rho_0\left(\frac{a}{a_0}\right)^{-3(1+3\omega)}+\frac{n}{\phi_0^2}\left(\frac{a}{a_0}\right)^{\frac{-2\delta}{n}},\\
\\
\frac{\ddot{a}}{a}=-\frac{4\pi G_0}{3}\rho_0(1+3\omega)\left(\frac{a}{a_0}\right)^{-3(1+3\omega)}+\frac{n}{\phi_0^2}\left(1-\frac{\delta}{n}\right)\left(\frac{a}{a_0}\right)^{\frac{-2\delta}{n}}.
\end{array}
\end{eqnarray}
\subsection{The acceleration universe}
For a rough estimate consider a spatially flat brane composed of cold dark
matter. In this case, the deceleration parameter reads
\begin{eqnarray}\label{2-15}
q:=-\frac{\ddot a}{aH^2}=\frac{1}{2}\Omega_m-(1-\frac{\delta}{n})\Omega_{(DE)},
\end{eqnarray}
where
\begin{eqnarray}\label{2-16}
\begin{array}{cc}
\Omega_m:=\frac{8\pi G_N\rho_m}{3H^2},\\
\\
\Omega_{(DE)}:=\frac{n}{H^2\phi^2}.
\end{array}
\end{eqnarray}
The critical case $q_0=0$ $(\delta/n=-(1-3\Omega^{(0)}_{(DE)})/2\Omega^{(0)}_{(DE)})$, describes
a coasting cosmology instead of being supported by K-matter \cite{Kolb}. It is also interesting that even negative values of $q_0$ are allowed for dust filled Universe,
since the $q_0<0$ case can be satisfied if $\delta/n>-(1-3\Omega^{(0)}_{(DE)})/2\Omega^{(0)}_{(DE)}$
which is in line with recent  measurements using Type Ia supernovae.
The deceleration parameter as a function of the redshift for some
selected values of the number of the extra dimensions, $n$, are displayed
in Fig. 1.  It is of some interest to determine the
redshift of the deceleration-acceleration transition predicted to
exist in our model. If we set $\delta=0.01$ (see equation (\ref{m5})),
 the transition redshift $(q(z_t) = 0)$ from cosmic deceleration $(q > 0)$ to acceleration $(q < 0)$ for a $11D$ bulk space will be $z_t\approx0.65$
which is in good agreement with the recent observations \cite{z}.
\begin{figure}
\centering
\includegraphics[width=8cm]{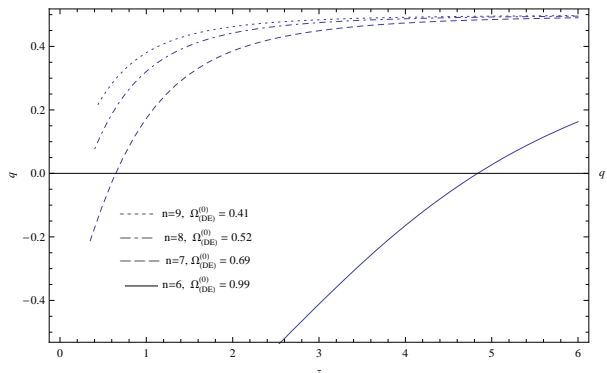}
\caption{Deceleration parameter as a function of the redshift for some selected values of the number of  extra dimensions $n$, and $\delta=0.01$. The value of the density parameter is obtained from table (I).}\label{Fig1}
\end{figure}
Also the lookback time, $\Delta t=t_0-t$, in this model (flat case) will be
\begin{eqnarray}\label{2-17}
\Delta t=
\frac{1}{H_0}\int_{0}^z\frac{1}{\left[(1-\Omega^{(0)}_{DE})(1+z')^5+\Omega^{(0)}_{DE}(1+z')^{2(\delta/n+1)}\right]^{\frac{1}{2}}}dz',
\end{eqnarray}
which shows the difference between the age of the Universe at the present
epoch and when a light ray at redshift $z$ was emmited.
Choosing $z=\infty$ in (\ref{2-17}) we have the present age of the Universe.
The lookback time curves as a function of the redshift for some
selected values of the number of the extra dimensions, $n$, are displayed
in Fig. 2.
\begin{figure}
\centering
\includegraphics[width=8cm]{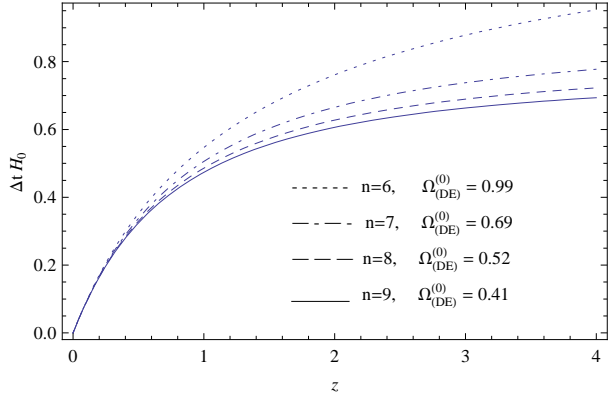}\\
\caption{Lookback time as a function of the redshift for some selected values
of the number of  extra dimensions $n$, and $\delta=0.01$. The lookback time decreases for higher values of $n$. The value of density parameter is obtained
from table (I).}\label{Fig2}
\end{figure}

Note that the geometric term $Q_{\mu\nu}$ obtained in (\ref{2-4}) can be rewritten
as
\begin{eqnarray}\label{2-21}
Q_{\alpha\beta}=\frac{2nh}{\phi^2H}u_\alpha u_\beta-\frac{n}{\phi^2}\left(3-2\frac{h}{H}\right)g_{\alpha\beta}.
\end{eqnarray}
One can define the corresponding  perfect fluid energy-momentum tensor
with geometric origin as
\begin{eqnarray}\label{2-22}
\begin{array}{cc}
Q_{\alpha\beta}:=-8\pi G\left[(\rho_{(DE)}+p_{(DE)})u_\alpha u_\beta+p_{(DE)}g_{\alpha\beta}\right],\\
\\
\rho_{(DE)}=\frac{3n}{8\pi G_0\phi_0^2}\left(\frac{a}{a_0}\right)^{-\delta(2/n-1)},\\
\\
p_{(DE)}=-\frac{3n-2\delta}{8\pi G_0\phi_0^2}\left(\frac{a}{a_0}\right)^{-\delta(2/n-1)},
\end{array}
\end{eqnarray}
where $\rho_{(DE)}$ and $p_{(DE)}$ are  the energy density and the pressure of the
extrinsic dark fluid respectively, with the equation of state parameter $\omega_{(DE)}$
\begin{eqnarray}\label{2-23}
\omega_{(DE)}:=\frac{p_{(DE)}}{\rho_{(DE)}}=-(1-\frac{2\delta}{3n}).
\end{eqnarray}
Therefore, it seems that from the point of view of a $4D$ observer, the effect of the extrinsic shape of a brane looks like  a perfect fluid with negative pressure.

The current value of the normal curvature radii of the FLRW Universe using (\ref{1-L})
will be
\begin{eqnarray}\label{m1}
L_0=\frac{n}{4(n+1)}K^{(ad)}\frac{g_i}{4\pi}^2l^3M_{Pl}^2|_{t=t_0},
\end{eqnarray}
which by inserting into  (\ref{2-16}) gives the current value of the density
parameter of  dark energy as
\begin{eqnarray}\label{m2}
\Omega^{(0)}_{DE}=\left(\frac{16\pi(n+2)}{\sqrt{n}l^3M_{Pl}^2K^{(ad)}g_i^2H}\right)^2\mid_{t=t_0}.
\end{eqnarray}
The width of the brane $l$, can be taken as a fundamental length related to the
minimum size  of the confined
standard model interactions, where  in the present cosmological settings, should be taken as localized
 baryonic matter (nucleons).  In the strong interactions physics, it is widely believed
that coupling to the lowest mass mesons determines the localization size of hadrons to be $1/m_{{meson}}$ in the manner first described by Yokawa
\cite{Yokawa}.
Further evidence in support of the idea that all hadrons, e.g. nucleons, couple strongly
to low mass mesons is the fact that pions are the principal secondary component
in high-energy collisions \cite{Bushnin}. Also, as we know, the strong coupling
of hadrons to light mesons is a general phenomena. These considerations,
together with Gell-Mann's totalitarian principle lead
Bahcall and Frantschi  to the concept of ``hadron barrier'': The minimum
size of any system containing a nucleon cannot then be appreciably less than
the range of the strong interaction of pions with nucleons i.e. $1/m_\pi$
\cite{Bahcall}.
Therefore, the minimum measurable length over
which matter fields  can be localized on the brane is  of the order of their Compton wavelength of pion \cite{naka}. Also it is close to the radius of
confinement, holding gluons and quarks inside hadrons \cite{okun}. Hence one may
 assume the width of the brane is of order of the Compton wavelength of the pion $m_{\pi^0}^{-1}=1.46\times10^{-13}\hspace{.2cm}\mathrm{cm}$,
or in a better approximation, the overlap of the quark and antiquark wave-functions in the pion, called the decay constant $f_{\pi^-}=130.41\hspace{.2cm} \mathrm{MeV}$
\cite{quark} (all hadrons have a size $\gtrsim1/f_{\pi^-}$). Therefore we set
\begin{eqnarray}\label{pion}
l\approx\frac{1}{f_{\pi^-}}=1.513\times10^{-13}\hspace{.2cm}\mathrm{cm},
\end{eqnarray}
 and $g_i^2/4\pi=16.8$, the strong interaction pion-nucleon coupling ``constant''
 \cite{coupling}.
 Also, the best fit from the Planck temperature data with Planck lensing gives  $H_0\approx\ 64.14$ $\mathrm{Km s}^{-1}\mathrm{Mpc}^{-1}$  \cite{Planck}. Hence the normal curvature radii  at the present epoch will be
\begin{eqnarray}\label{m3}
L_0=\frac{nK^{(ad)}g_i^2M_{Pl}^2}{16\pi(n+2)f^3_{\pi^-}}\mid_{t=t_0}\approx\frac{n(n-2)}{n+2}11.44\times10^{27}\mathrm{cm}.
\end{eqnarray}
In 1921,  Kaluza suggested to unify the gravitational and electromagnetic interaction within the framework of a five multidimensional gravity theory. The fifth coordinate was made
invisible through  the cylinder condition: it was assumed that in the fifth direction, the world curled up into a cylinder of very small radius (Planck's length). In modern extension of KK ideas, the extra dimensions can be large and sometimes non-compact and the typical length of the extra dimensions will be of order  $M_{Pl}^{\frac{2}{n}}/M_{D}^{\frac{2}{n}+1}$ \cite{Arkani}. For $n=1$, the radius of internal space will be of order
$10^{13}$ cm.  This is approximately the diameter of the solar system, quite unsuitable as an internal dimension. On the other hand if $n = 2$, we obtain the sub-millimeter size for the internal space. According to (\ref{m3}),
the normal radii (a reasonable distance in the bulk space for gravity) of brane in our model is of order of the Hubble distance, like Deffayet, Dvali and Gabadadze
model \cite{3}. Also the density parameter of dark energy becomes
\begin{eqnarray}\label{m4}
\Omega^{(0)}_{(DE)}=\left(\frac{16\pi(n+2)f^3_{\pi^-}}{\sqrt{n}M_{Pl}^2K^{(ad)}g_i^2H}\right)^2\mid_{t=t_0}\approx\frac{1.221}{n}\left(\frac{n+2}{n-2}\right)^2.
\end{eqnarray}
Besides the BBN bound, namely
\begin{eqnarray}\label{m5}
|\frac{\dot{G}_N}{G_N}|_{t=t_0}<0.01H_0,
\end{eqnarray}
is also comparable with inequalities (\ref{s1})-(\ref{2-12}) help us to estimate $\delta\sim0.01$.

The energy density of dark energy at the present epoch will be
\begin{eqnarray}\label{an}
\begin{array}{cc}

\rho_{(DE)}^{(0)}=\frac{3n}{8\pi}\frac{M_{Pl}}{l_{Pl}L^2}|_{t=t_0}\approx\frac{3}{2\pi}\left(\frac{n+2}{\sqrt{n}(n-2)}\right)^2\left(\frac{4\pi}{g_i^2}\right)^2G_Nm_{\pi^0}^6\mid_{t=t_{0}}.
\end{array}
\end{eqnarray}
Let us point out that, nearly half a century ago, an interpretation
of the cosmological constant through vacuum energy
density was pioneered by Zeldovich \cite{Zeldovich}. He conjectured
\begin{eqnarray}\label{zel}
\rho_{(DE)}=\frac{\Lambda}{8\pi G_N}\sim G_Nm^6,
\end{eqnarray}
where $\Lambda$ is the cosmological constant and $m$ is close to the pion
mass. Note that, the Zeldovich conjecture predicts a constant vacuum energy density, while
equation (\ref{an}) (which is an amendment to his conjecture)  indicates a dark energy density
which is not constant but changes with the expansion of the
Universe.

 In the following table, we find the density parameter of dark energy,  age of the Universe and state parameter of dark energy for some specific number of extra dimensions.

\begin{center}{Table (I)}\\
\begin{tabular}{|c|c|c|c|c|c|}
\hline  n& $\Omega^{(0)}_{(DE)}$ &Age of Universe
$(t_0)$&$\omega_{(DE)}$\\
\hline 6 & 0.99 &28.80 Gyr&-0.9989\\
\hline 7 & 0.69 & 13.73 Gyr&-0.9990\\
\hline 8 & 0.52 & 12.09 Gyr&-0.9992\\
\hline 9 & 0.41 & 11.35 Gyr&-0.9993\\
\hline
\end{tabular}
\end{center}
\vspace{.5cm}\noindent\\
As one can see in  table (I), the  case of $n=7$ ($11D$
bulk space) is particularly interesting. The contribution of baryonic and cold
dark matters is nothing for  $2\leq n\leq 6$, $(\Omega^{(0)}_{(DE)} \geq1)$. On the other hand, observations do not agree with the theoretically
obtained value of the age of Universe and density parameter for $n>7$. While for $n=7$, not only the density fraction,  age of the Universe and the equation of state parameter are in good agreement with observations and results of
the Planck temperature data with Planck lensing \cite{Planck}, but
also  they have the maximum possible values. The spacial case of $11D$ ambient
space not only is in agreement with observations but also has deeper geometrical
meaning in our model. As was mentioned in the third item of introduction,
one of our goals in this paper was to find a geometrical mechanism for unification
of fundamental forces, using the group of isometries of the non-compact ambient
space. As shown in \cite{Maraner},  we need at least seven non-compact extra dimensions
to realize the standard Model group $U(1)\times SU(2)\times SU(3)$. The question of the true dimensionality of the world is surely one of the deepest questions
that one can possibly ask in physics. Strong motivation for considering space as multidimensional comes from theories which incorporate gravity in a reliable manner; � string theory and M-theory. As we know,
to facilitate the construction of supersymmetry and supergravity Lagrangians, we need
$11D$  spacetime which is now recognized as the low energy effective description
of $11D$ M-theory \cite{W}. Also to construct consistent superstring theories, the requirement is a $10D$ spacetime. M-theory is an $11D$ extension of superstring
theory, where proponents believe that the $11D$ theory unifies all five $10D$
string theories. The idea is  the unique supersymmetric theory in $11D$, with its low-entropy matter content and interactions fully determined, and can be obtained as the strong coupling limit of type IIA string theory because a new dimension of space emerges as the coupling constant increases \cite{Schwarz}.
As we know, the various superstring theories are related by dualities which
allow the description of an object in one string model to be related to the
description of a different object in another one. These dualities imply that
each of string theories is a different manifestation of a single underlying
$11D$
M-theory. All in all, one may conclude that perhaps the real ``home'' of
string theory is not $10D$, but possibly $11D$ M-theory which could be reduced
down to $11D$ supergravity as well as $10D$ string theory.

 \section{Concluding remarks}
  We have derived the covariant gravitational equations of a brane world model embedded isometrically in a bulk space with arbitrary number of dimensions.
  The extrinsic shape of the brane has a fundamental role in our
approach. It makes a similarity between KK gravity and modern embedding
program. Also the induced $4D$ gravitational constant  depends on the  extrinsic
shape of our Universe.
The gravitational model thus obtained was then applied to cosmology. Assuming the FLRW spacetime embedded isometrically in a bulk space with arbitrary spacelike extra dimensions, we have re-derived the generalized Friedmann equations. The obtained field
equations contain the geometric fluid with its properties  related to
the extrinsic shape of the brane via the gravitational ``constant''. Hence, from point of view of a $4D$ confined observer, (which does not have access to the extra
dimensions), the late time acceleration of the Universe looks like as the effect of  some phenomenological dark fluid.

\section{Acknowledgements}
We thank H. R. Sepangi and M. D. Maia for comments and suggestions.


\end{document}